\documentclass[fleqn,usenatbib]{mnras}

\usepackage{newtxtext,newtxmath}
\usepackage[T1]{fontenc}
\usepackage{ae,aecompl}

\usepackage{graphicx}	% Including figure files
\usepackage{amsmath}	% Advanced maths commands
\usepackage{amssymb}	% Extra maths symbols

\usepackage{color} 

\newcommand{\Msun}{\mbox{$\mathrm{M}_{\odot}$}}

\newcommand{\Porb}{\mbox{$P_{\mathrm{orb}}$}}

\def\maa{$\alpha \alpha$}
\def\mga{$\gamma \alpha$}

\title[The double-degenerate progenitors of SNIa]{Where are the
  double-degenerate progenitors of type Ia supernovae?}

\author[Rebassa-Mansergas et al.]{
A. Rebassa-Mansergas$^{1,2}$\thanks{E-mail: alberto.rebassa@upc.edu},
S. Toonen$^{3}$, V. Korol$^{4}$,
S. Torres$^{1,2}$
\\
$^{1}$ Departament de F\'\i sica, Universitat Polit\`ecnica de
Catalunya, c/Esteve Terrades 5, 08860 Castelldefels, Spain\\
$^{2}$ Institute for Space Studies of Catalonia, c/Gran Capit\`a 2--4,
Edif. Nexus 201, 08034 Barcelona, Spain\\
$^{3}$ Anton Pannekoek Institute for Astronomy, University of
Amsterdam, 1090 GE Amsterdam, The Netherlands\\
$^{4}$ Leiden Observatory, Leiden University, PO Box 9513, 2300 RA,
Leiden, the Netherlands
}

\date{Accepted XXX. Received YYY; in original form ZZZ}

\pubyear{2018}

\begin{document}
\label{firstpage}
\pagerange{\pageref{firstpage}--\pageref{lastpage}}
\maketitle

\begin{abstract}
Double white  dwarf binaries with  merger timescales smaller  than the
Hubble time and  with a total mass near the  Chandrasekhar limit (i.e.
classical Chandrasekhar population) or  with high-mass primaries (i.e.
sub-Chandrasekhar population)  are potential supernova type  Ia (SNIa)
progenitors.   However, we  have not  yet unambiguously  confirmed the
existence of these objects observationally, a fact that has been often
used to criticise  the relevance of double white  dwarfs for producing
SNIa.   We  analyse  whether  this   lack  of  detections  is  due  to
observational effects. To that end  we simulate the double white dwarf
binary population in  the Galaxy and obtain synthetic  spectra for the
SNIa progenitors.  We demonstrate  that their identification, based on
the detection of H$\alpha$ double-lined  profiles arising from the two
white dwarfs in the synthetic spectra, is extremely challenging due to
their  intrinsic faintness.   This  translates  into an  observational
probability  of finding  double white  dwarf SNIa  progenitors in  the
Galaxy  of  $(2.1\pm1.0)\times10^{-5}$ and  $(0.8\pm0.4)\times10^{-5}$
for the  classical Chandrasekhar and the  sub-Chandrasekhar progenitor
populations,   respectively.   Eclipsing   double  white   dwarf  SNIa
progenitors  are   found  to   suffer  from  the   same  observational
effect. The next generation  of large-aperture telescopes are expected
to help in  increasing the probability for detection  by $\sim$1 order
of magnitude.  However, it is  only with forthcoming observations such
as those provided  by LISA that we expect to  unambiguously confirm or
disprove the existence  of double white dwarf SNIa  progenitors and to
test their importance for producing SNIa.
\end{abstract}

\begin{keywords}
(stars:) white dwarfs; (stars:) binaries: spectroscopic; (stars:)
  supernovae: general
\end{keywords}

%%%%%%%%%%%%%%%%%%%%%%%%%%%%%%%%%%%%%%%%%%%%%%%%%%

%%%%%%%%%%%%%%%%% BODY OF PAPER %%%%%%%%%%%%%%%%%%

\section{Introduction}

Supernovae Type Ia  (SNIa) are one of the most  luminous events in the
Universe, which makes them ideal  tools for cosmological studies since
they can be detected at very large distances. In particular, SNIa have
been  used to  prove  the  accelerated expansion  of  the Universe,  a
discovery  which  was awarded  the  Nobel  prize  in physics  in  2011
\citep[e.g.][]{Riess98, Perlmutter99, AstierPain12}. However, there is
not  yet a  consensus on  the  leading paths  to SNIa  \citep[see][for
  recent reviews]{Liv18,  Sok18, Wang18}. This  progenitor uncertainty
may  introduce   some  not   yet  known   systematic  errors   in  the
determination of extragalactic distances, thus compromising the use of
SNIa as standard candles \citep{Linden09, Howell2011}.

Several evolutionary channels  have been proposed that lead  to a SNIa
explosion.    For   a    comprehensive   review,   see   \citet{Liv18,
  Wang18}. Among  these, the two  classical scenarios are  the single-
and the double-degenerate channels. In the single-degenerate channel a
WD in a binary system accretes  mass from a non-degenerate donor until
it grows  near the  Chandrasekhar limit  \citep{WhelanIben73, Han2004,
  nomoto2018}. In  the double-degenerate  channel two  WDs in  a close
binary  system  merge due  to  angular  momentum  loss caused  by  the
emission of  gravitational waves and  the resulting merger has  a mass
near  the  Chandrasekhar  limit  \citep{WhelanIben73,  Iben+Tutukov84,
  Liu2018}.   Additional evolutionary  channels for  SNIa include  the
double-detonation mechanism  \citep{Woosley1986, Livne1995, Shen2012},
the  violent merger  model \citep{Pak10,  Sat16}, the  core-degenerate
channel  \citep{Sparks1974,  Liv2003,  KashiSoker11, Wang2017}  and  a
mechanism  which  involves  the  collision of  two  WDs  \citep{Ben89,
  Kushnir2013, Aznarsiguan13}. In the  double-detonation scenario a WD
accumulates helium-rich  material on its surface,  which is compressed
and ultimately detonates.  The compression wave propagates towards the
center of the WD and a second detonation occurs near the center of its
carbon-oxygen core.   In the violent  merger model, the  detonation of
the  white dwarf  core is  initiated during  the early  stages of  the
merger.  This can happen, for example, due to compressional heating by
accretion  from  the  disrupted  secondary  or  due  to  a  preceeding
detonation of  accreted helium (alike the  double-detonation scenario)
that  is  ignited  dynamically   \citep{Pak10,  Pak11,  Pak12,  Pak13,
  Guil2010, Kas15,  Sat15, Sat16}.  In the  core-degenerate scenario a
WD merges with the hot core  of an asymptotic giant branch star during
(or after)  a common envelope  phase. Finally, the  evolutionary phase
involving  the collision  of two  WDs requires  a tertiary  star which
brings the  two WDs to  collide due  to the Kozai-Lidov  mechanism, or
dynamical interactions in  a dense stellar system, where  this kind of
interaction is more likely to happen.

The viability of the above  described SNIa formation channels has been
intensively studied  during the last several  years both theoretically
and   observationally   --   see,   for  example,   the   reviews   by
\citet{Hil2013,    Maoz2014,    Wang18,    Sok18}    and    references
therein. However, there is not yet an agreement on how these different
evolutionary paths contribute to the observed population of SNIa, with
all channels presenting advantages and drawbacks.  In particular, from
the theoretical perspective, it is not clear whether double WD mergers
arising from the double-degenerate channel  result in a SNIa explosion
or  rather  in  an  accretion-induced   collapse  to  a  neutron  star
\citep{NomotoIben1985,  Shen2012}.   The  hypothesis that  WD  mergers
containing   less    massive   primaries,   i.e.    the    so   called
sub-Chandrasekhar  WDs,  play  a  decisive  role  in  reproducing  the
observed   SNIa    luminosity   function   is   also    under   debate
\citep[e.g.][]{Shen17}.    It   is   also   fair   to   mention   that
double-degenerate models predict a delay time distribution which is in
better   agreement   with   the    one   derived   from   observations
\citep[e.g.][]{Maoz2017b}.      Furthermore,    several     additional
observational analyses have provided support for the double-degenerate
channel   \citep{Tovmassian2010,    Rodriguez-Gil2010,   Gonzalez2012,
  Olling2015}.   However, perhaps  with the  exception of  the central
binary    system    of    the   planetary    nebula    Henize    2-428
\citep{Santander2015}, there is no single system yet that has robustly
been confirmed  as a double-degenerate SN\,Ia  progenitor.  The nature
of Henize 2-428 as a direct SNIa double-degenerate progenitor has been
criticised by \cite{Garciaberro2016}, who claim that the binary system
may be formed by  a WD and a low-mass main  sequence companion, or two
WDs   of    smaller   combined    mass   than   that    estimated   by
\citet{Santander2015}.

Finding close double-degenerate binaries  is not straightforward since
their spectra are  virtually identical to those of  single WDs. Hence,
their identification has been mainly  based on the detection of radial
velocity   variations    \citep{Marsh95,   Maxtedetal2000,   Maxted02,
  MaxtedMarsh99,  Brown2013,  Brown2016,  Kilic2017,  Rebassa17}.   In
particular,  the observational  effort  carried out  by  the ESO  SNIa
Progenitor   (SPY)  Survey   \citep{Napiwotzki01,  Napiwotzki07}   has
provided radial velocities  for hundreds of double  WDs, including the
identification       of       several      double-lined       binaries
\citep{Koester01}. More recently, \citet{Breedt2017} analysed multiple
spectra available for individual WDs  in the SDSS to preselect targets
displaying variability for follow-up  observations. Although no direct
SNIa progenitors  has been  identified, the  analysis of  both samples
(SPY and SDSS)  have allowed constraining the  binary fraction, merger
rate  and  separation  distributions  of  double  WDs  in  the  Galaxy
\citep{Maoz2012, BadenesMaoz12,  Maoz2018} as well as  identifying hot
sub-dwarf plus white dwarf  binaries \citep{Geier2010} and white dwarf
plus M dwarf binaries \citep{Rebassa2011, Rebassa2016}.

The  fact that  not  a single  double-degenerate  progenitor has  been
unambiguously identified  among our currently available  large samples
of  double  WDs  may  be  used   as  an  argument  indicative  of  the
double-degenerate  mechanism  not being  a  viable  channel for  SNIa.
However, it is also fair  to mention that identifying SNIa progenitors
not only requires measuring the orbital periods but also the component
masses of the  two WDs. Even with such large  superb samples of double
WDs at  hand, only  few of  them have  well measured  component masses
\citep[see][and reference therein]{Rebassa17}. The obvious question is
then: what  is the  probability of identifying  SNIa double-degenerate
progenitors?    Or   in   other   words:  are   we   not   identifying
double-degenerate   progenitors   because    it   is   observationally
challenging  or because  they simply  do not  exist?  We  assess these
questions quantitatively in  this paper.  To that end  we simulate the
close double WD population in the Galaxy and we analyse whether or not
the SNIa  progenitors in  our simulations  would be  easily identified
observationally with our current telescopes and instrumentation.

\begin{table}
\caption{The  total  number of  double  WDs  in the  four  populations
  considered in  this work. The  numbers vary according to  the common
  envelope  prescription adopted  in  the simulations.  Note that  our
  simulations exclude all  binaries in which any of  the WD components
  has a $g$ magnitude $>23$ mag.}
\begin{center}
\begin{tabular}{ccccc}
\hline
\hline
CE formalism & Ch. direct & SCh. direct & Ch. nmer & SCh. nmer \\
\hline
$\alpha\alpha$  & 176        & 51         &  14065   &    7431  \\
$\gamma\alpha$  & 107        & 22         &  21596   &    8476  \\
\hline
\end{tabular}
\end{center}
\label{t-samples}
\end{table}

\begin{figure*}
    \includegraphics[width=0.9\textwidth]{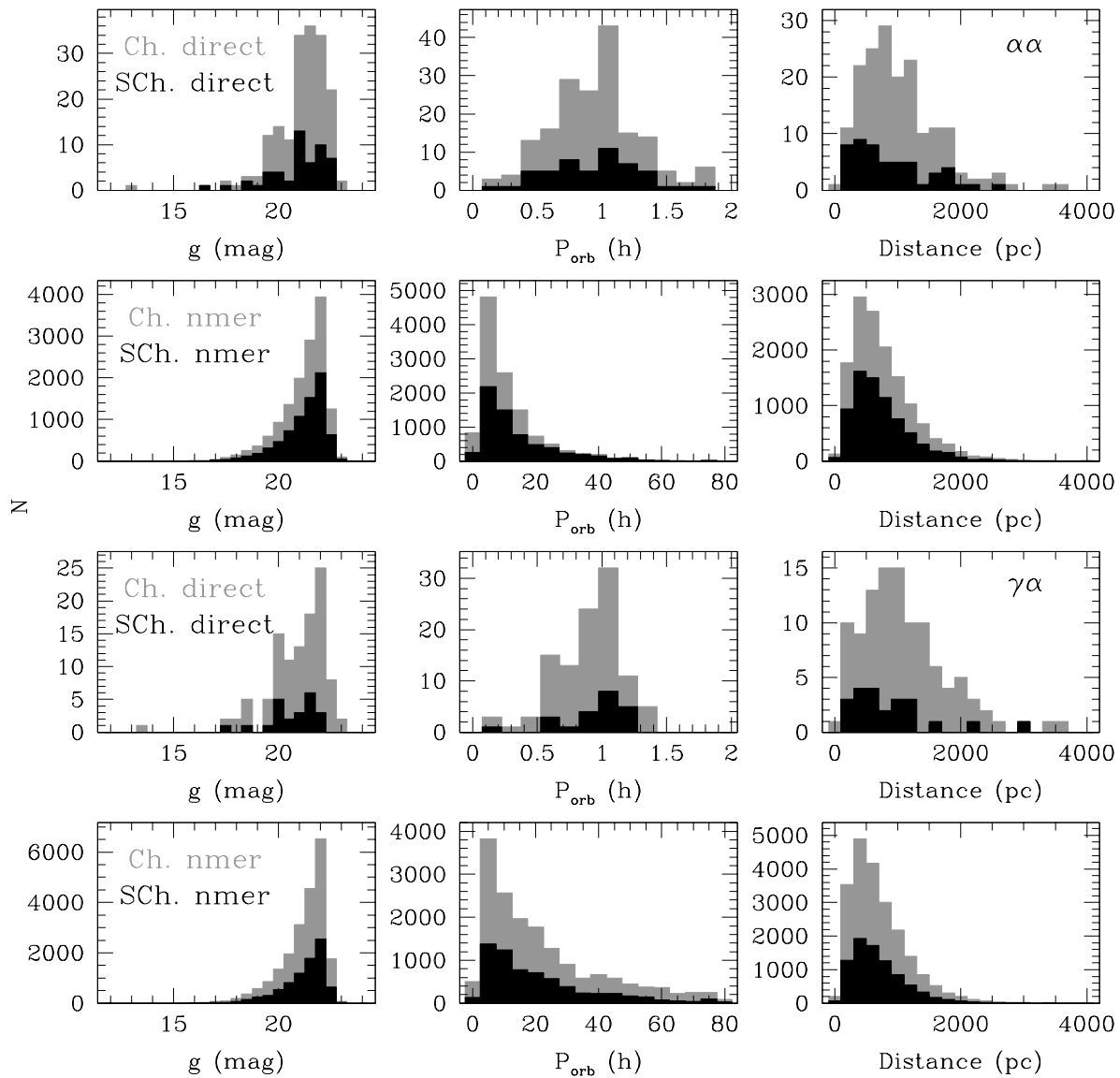}
    \caption{The distribution  of $g$ magnitudes, orbital  periods and
      distances  for  the four  populations  considered  in this  work
      (black for SCh.  and gray for Ch. progenitors) when a cut off at
      $g=23$   mag   is   adopted   for   the   WD   components   (see
      Table\,\ref{t-samples}).    The   top  and   middle-top   panels
      illustrate  systems that  evolved through  $\alpha\alpha$ common
      envelopes,  the bottom-middle  and  bottom  panels systems  that
      evolved through $\gamma\alpha$ common envelopes.}
    \label{f-param}
\end{figure*}

\section{Synthetic binary population models}
\label{s-models}

We create synthetic  models for the Galactic population  of double WDs
by means  of the binary  population synthesis (BPS) method.  We employ
the  code  \texttt{SeBa}  \citep{Por96,Too12,Too13}  to  simulate  the
formation and evolution of interacting binaries producing double WDs.

The initial binaries are generated according to a classical set-up for
BPS calculations in the following way:
\begin{itemize}
\item We  draw a mass from  the initial mass function  of \cite{Kro93}
  within the range 0.1-100\Msun;
\item The  masses of the companion  stars follow a uniform  mass ratio
  distribution between 0 and 1 \citep{Rag10,Duc13,Ros14,Cojocaru17};
\item The orbital separation $a$  is drawn from a uniform distribution
  in $\rm  log(a)$ \citep{Abt83}. Note that  a log-normal distribution
  peaking  at   around  $10^5$d   is  preferred   observationally  for
  Solar-type stars \citep{Duq91,Rag10,  Duc13,Moe17}. This affects the
  number of simulated double WDs to less than 5\% \citep{Too17};
\item  The   eccentricities  ($e$)   follow  a   thermal  distribution
  \citep{Heg75}: $f(e) = 2e$ with $0<e<1$;
\item We adopt a constant binary fraction of 50\% which is appropriate
  for  A-, F-,  and G-type  stars \citep{Rag10,Duc13,Ros14,Moe17}.   A
  binary fraction of 75\% \citep[as  observed for O- and B-type stars,
    e.g.][]{San12} would increase the numbers of double WDs by 36\%;
\item The orbital inclinations $i$ are obtained from a uniform
  distribution of $\sin i$.
\end{itemize}

It  was shown  in \cite{Too14}  that the  main sources  of differences
between the  synthetic models  of different  BPS codes  is due  to the
choice of  input physics and  initial conditions. For double  WDs, the
most  impactful assumption  is that  of the  physics of  unstable mass
transfer; in which  systems is the mass  transfer not self-regulating,
and what  is the effect  on the  binary orbit and  stellar components?
Unstable mass transfer gives rise to a short phase in the evolution of
a binary system in which both stars share a common-envelope (CE). Even
though  CE-evolution  plays an  essential  role  in the  formation  of
compact binaries,  and despite the  enormous effort of  the community,
the    CE-phase    is    poorly    understood    \citep[e.g.][for    a
  review]{Iva13}. For this  reason we employ two  different models for
the CE-phase, model \maa\, and model \mga, which are described below.

The  classical  model (model  \maa)  is  based  on the  energy  budget
\citep{Pac76, Tut79, Web84, Liv88}:
\begin{equation}
E_{\rm gr} = \alpha (E_{\rm orb,initial}-E_{\rm orb,final}),
\label{eq:alpha-ce}
\end{equation}
where $E_{\rm gr}$ is the binding energy of the envelope mass, $E_{\rm
  orb}$ is the orbital energy,  and $\alpha$ the efficiency with which
orbital energy  is consumed to  unbind the CE. We  approximate $E_{\rm
  gr}$ by:
\begin{equation}
E_{\rm gr} = \frac{GM M_{\rm env}}{\lambda R},
\label{eq:Egr}
\end{equation} 
where $M$  is the mass of  the donor star, $M_{\rm  env}$ its envelope
mass, $\lambda$ the  envelope structure parameter, and  $R$ the radius
of  the donor  star. Here  we  adopt $\alpha\lambda=2$  as derived  by
\cite{Nel00} by  reconstructing the formation  of the second WD  for a
sample of observed double WDs.

\begin{figure*}
    \includegraphics[width=0.7\textwidth]{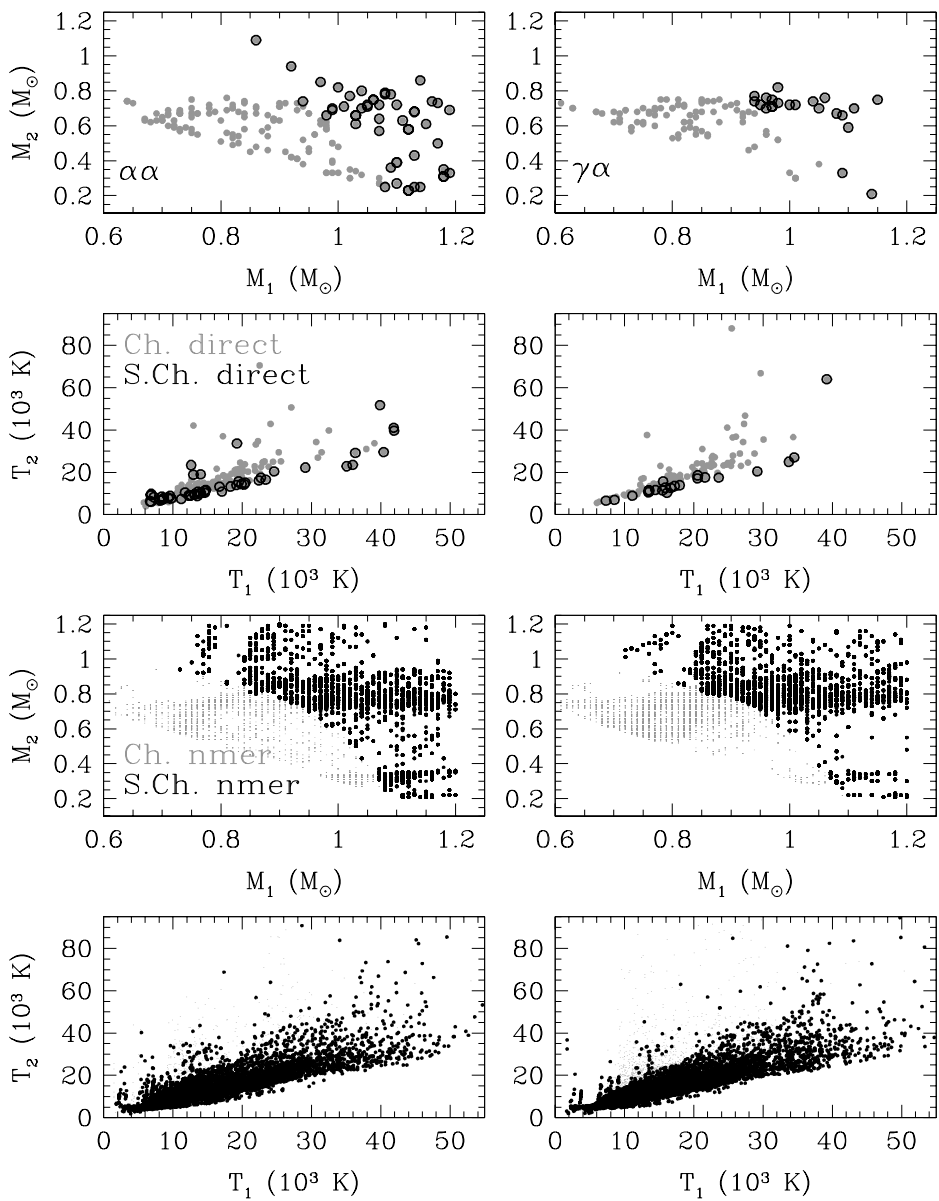}
    \caption{Comparison  between the  component  masses and  effective
      temperatures of the two WDs  for the four populations considered
      in  this work  (see  Table\,\ref{t-samples}).   The left  panels
      illustrate  systems that  evolved through  $\alpha\alpha$ common
      envelopes,  the  right  panels   systems  that  evolved  through
      $\gamma\alpha$ common envelopes. Note  that the SCh. progenitors
      (black circles)  are a sub-population  of the Ch.  samples (gray
      circles).}
    \label{f-param2}
\end{figure*}

The alternative model  (model \mga) is inspired by the  (same) work of
\cite{Nel00}. In order  to explain the observed mass  ratios of double
WDs, \cite{Nel00} propose an  alternative CE-formalism, which is based
on the angular momentum budget:
\begin{equation}
\frac{J_{\rm initial}-J_{\rm final}}{J_{\rm initial}} = \gamma \frac{\Delta M}{M+ m},
\label{eq:gamma-ce}
\end{equation} 
where $J$ is the  angular momentum of the binary, and  $m$ the mass of
the  companion. We  adopt  $\gamma =  1.75$ \citep[see][]{Nel01}.   In
model  \mga\, when  a CE  develops, Eq.\,\ref{eq:gamma-ce}  is applied
unless the binary contains a compact  object or the CE is triggered by
the Darwin-Riemann  instability \citep{Dar1879, Hut1980}.   For double
WDs,    the   first    CE    is   typically    simulated   with    the
$\gamma$-parametrization, and the second with the $\alpha$-formalism.

\begin{figure*}
    \includegraphics[angle=-90,width=\columnwidth]{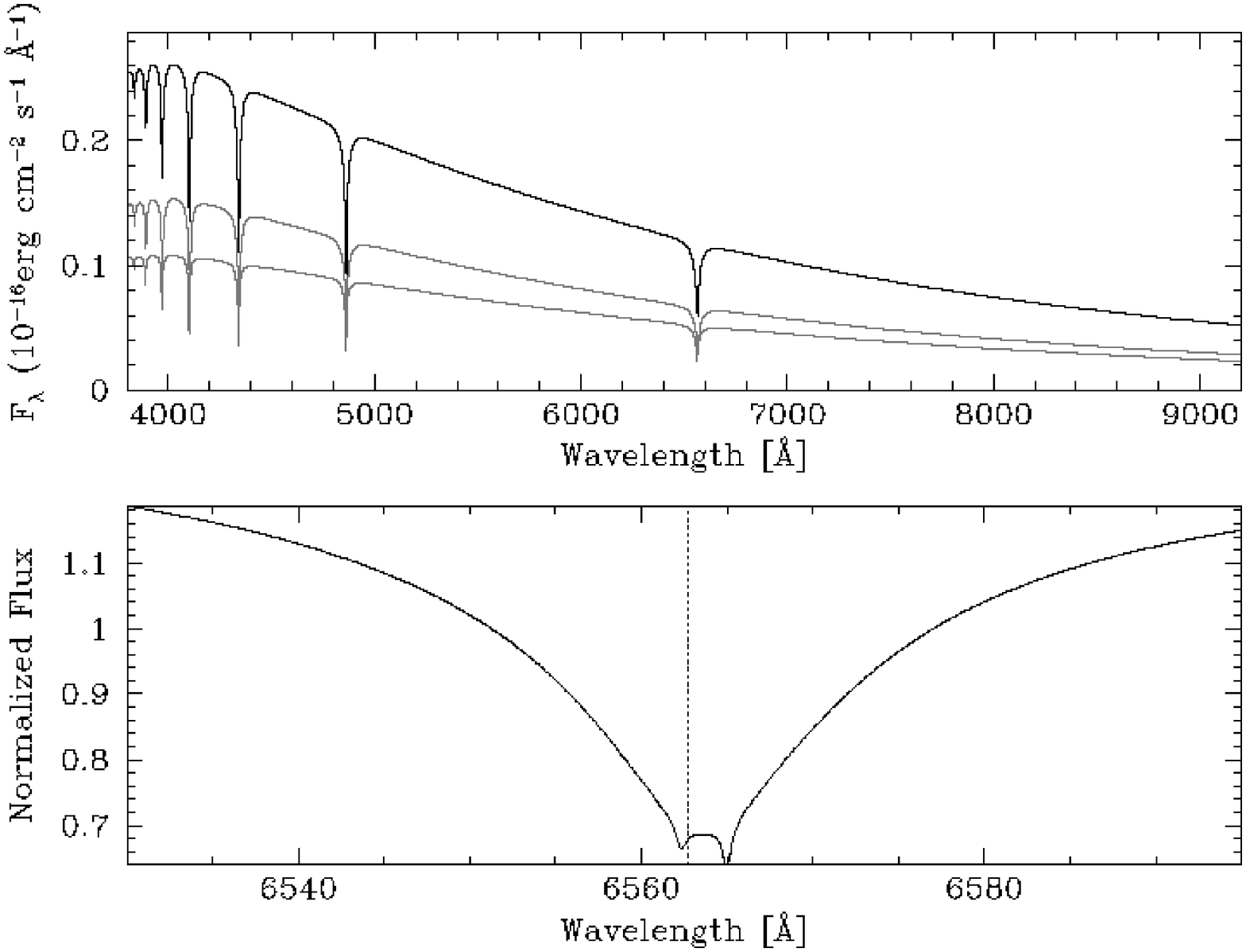}
    \includegraphics[angle=-90,width=\columnwidth]{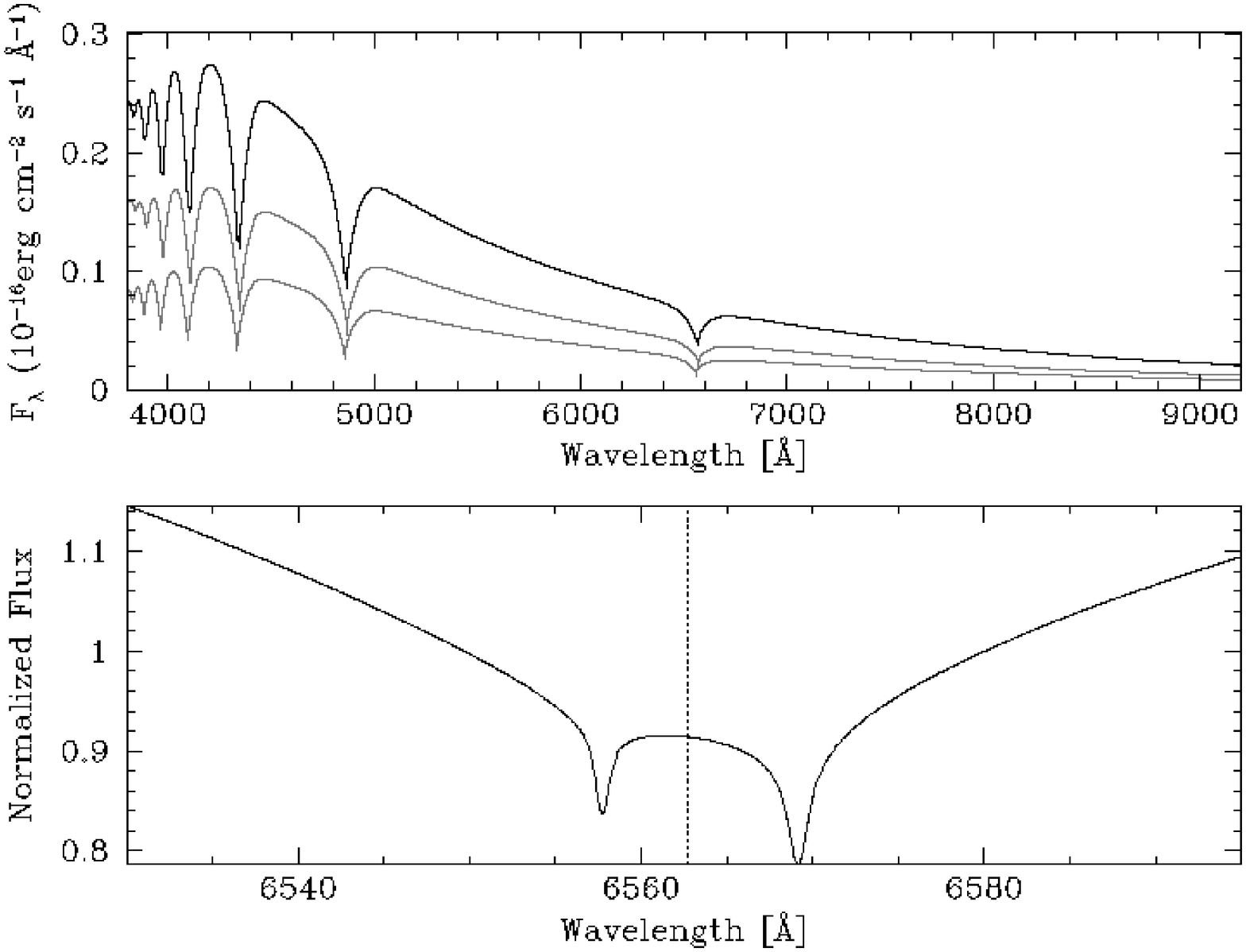}
    \caption{Top panels:  two examples of simulated  double WD spectra
      (black solid lines).  The individual  WD components are shown as
      gray solid  lines.  Bottom  panels: a  zoom-in to  the H$\alpha$
      line of  the combined spectra.   The temperatures and  masses of
      the      WD     components      are     $M_1=0.76\,$M$_{\odot}$,
      $M_2=0.67\,$M$_{\odot}$,   $T_1=8315$\,K,  $T_2=7715$\,K   (left
      panels)  and  $M_1=1.01\,$M$_{\odot}$,  $M_2=0.72\,$M$_{\odot}$,
      $T_1=17980$\,K, $T_2=13931$\,K (right panels).}
    \label{f-DD}
\end{figure*}

It has also  been proposed that the  first WD is not  formed through a
CE-phase,   but  through   stable,   non-conservative  mass   transfer
\citep{Woo12,  Pas12, Ge15}.  The  effect  on the  orbit  is a  modest
widening, similar  to that of  the $\gamma$-formalism. A BPS  study of
the  implications  on  the  double  WD  population  of  the  increased
stability of mass transfer, is beyond the scope of this paper.

To  study the  visibility  of the  double  WDs in  our  Milky Way,  we
convolve the BPS  data with the Galactic star  formation history (SFH)
and apply a WD cooling. The SFH is based on the model by \cite{Boi99},
which adopts a  total mass in stars  of 3.8$\times10^{10}$ M$_{\odot}$
and is a  function of both time  and position in the  Milky Way.  Full
details  on our  SFH model  can be  found in  \citet{Too13}, including
information  on   the  Galactic   components  adopted.    The  $ugriz$
magnitudes of the  WDs are estimated by their  distances, while taking
into account  extinction \citep{Schlegel1998} and cooling  through the
evolutionary  sequences. In  this work  we assume  all the  WDs to  be
composed of  pure hydrogen-rich atmospheres,  i.e.  DA WDs,  and hence
adopt  the cooling  sequences developed  for DA  WDs of  \citet{Hol06,
  Kow06,                    Tre11}\footnote{See                   also
  http://www.astro.umontreal.ca/bergeron/CoolingModels.}.  Knowing the
magnitudes of each WD component we can easily derive the magnitudes of
the  double WD  system by  summing up  the individual  fluxes in  each
band. This is a valid  assumption for close (unresolved) binaries such
us the progenitors  of SNIa.  For more details of  the Galactic model,
see \cite{Too13}.  Here,  we only consider systems where  at least one
component   has  a   g-band  magnitude   below  23   magnitudes  since
observations of  fainter systems  would be extremely  challenging.  We
also note that we only simulate the hydrogen-rich double WD population
in the Galaxy,  i.e. double DA WDs.   We define the primary  WD as the
first  formed WD,  the  secondary  is the  second  formed WD.   Hence,
hereafter all parameters associated with the primary and secondary WDs
will be  denoted by the  suffixes 1 and  2, respectively.  It  is also
important  to mention  that, once  the double-degenerate  binaries are
formed, we take into account  angular momentum losses by gravitational
wave  radiation, which  reduce  the orbital  separation until  present
time.

For  the present  work, and  based  on the  above described  numerical
simulations, we  define four  populations of interest:

\begin{itemize}
\item The  'Ch.  direct'  SNIa progenitor population,  which comprises
  double WDs that  merge within the Hubble time and  with a total mass
  exceeding 1.3\,\Msun\, (we  adopt this value as a  lower limit since
  SNIa explosions occur near the Chandrasekhar mass).
\item The 'SCh.  direct'  progenitor population, which includes double
  WDs that merge  within the Hubble time  leading to sub-Chandrasekhar
  explosions.  To select  these systems we apply the  condition $M_2 >
  -10.2041 \times (M_1  - 0.85)^2 + 0.805$ (or $M_1  > -10.2041 \times
  (M_2 - 0.85)^2  + 0.805$) provided by  \citet{Shen17}, which selects
  massive  primaries that  have  higher  gravitational potentials  and
  massive  secondaries that  yield more  directly impacting  accretion
  streams. These two processes make it more likely a sub-Chandrasekhar
  WD to explode.
\item The 'Ch.  nmer' SNIa progenitor population, which is the same as
  population 1  (Ch.  direct) but  for WD  binaries that do  not merge
  within the Hubble time.
\item The 'SCh.   nmer' population, which is the same  as population 2
  (SCh. direct)  but for systems that  do not merge within  the Hubble
  time.
\end{itemize}

As  we   will  show   in  Section\,\ref{s-results},   considering  the
non-merger samples  (i.e.  the Ch.   nmer and SCh.   nmer populations)
allows  deriving  more  sound   results  regarding  the  observational
properties of SNIa progenitors.   In Table\,\ref{t-samples} we provide
the number  of double WDs in  each population depending on  the common
envelope prescription used in our simulations.

The $g$ magnitude,  orbital period and distance  distributions as well
as  the   comparison  between  the  component   masses  and  effective
temperatures of  the two WDs  for the four considered  populations are
illustrated in Figure\,\ref{f-param}  and Figure\,\ref{f-param2}. From
the Figures  one can clearly  see that  the number of  non-merger SNIa
progenitors is  significantly larger than that  of direct progenitors,
and that  the orbital period  distributions for direct  and non-merger
systems are substantially different. 

\begin{figure}
    \includegraphics[width=\columnwidth]{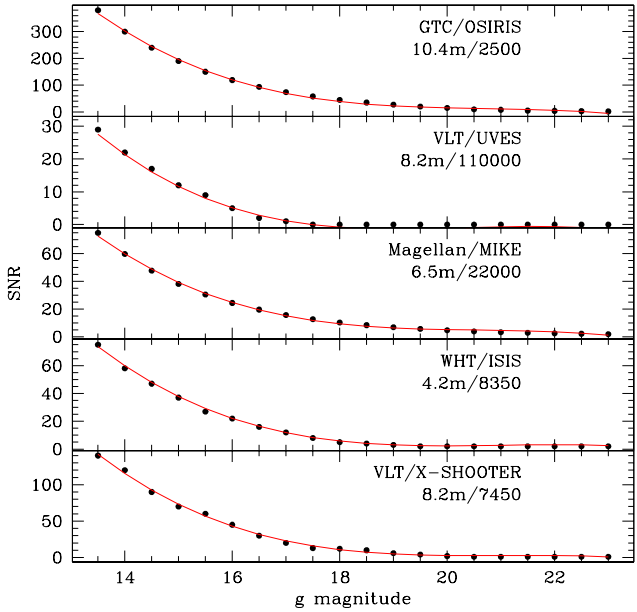}
    \caption{SNR  as  a  function   of  $g$  magnitude  (assuming  0.5
      magnitude bins) for the telescope/instrument pairs considered in
      this work,  and fixing  the exposure time  at 10  minutes (black
      solid dots). The red solid lines are third order polynomial fits
      to the data. The telescope apertures and resolving powers of the
      spectrographs are also indicated.}
    \label{f-snr}
\end{figure}

\section{The double-degenerate synthetic spectra}
\label{s-spectra}

The population  synthesis code described  in the previous  section has
provided us with masses, effective temperatures, surface gravities and
radii  of the  binary components,  as  well as  with orbital  periods,
orbital inclinations  and distances  to each  SNIa progenitor  in four
different populations.  Here we develop  a method for  obtaining their
synthetic spectra.

In a first  step we obtain a synthetic spectrum  for each WD component
by interpolating  the corresponding effective temperature  and surface
gravity  values on  an updated  grid  of model  atmosphere spectra  of
\citet{Koester10}.   The  grid  contains   612  spectra  of  effective
temperatures ranging from 6,000~K to 10,000~K in steps of 250\,K, from
10,000~K to 30,000~K in steps of 1,000~K, from 30,000~K to 70,000~K in
steps of 5,000~K and from 70,000~K to 100,000~K in steps of 10,000\,K,
and surface  gravities ranging  between 6.5 and  9.5\,dex in  steps of
0.25\,dex for  each effective  temperature. The model  spectra provide
the astrophysical fluxes at the  surface of the WDs ($F_\mathrm{wd}$),
which we convert into observed fluxes ($f_\mathrm{wd}$) using the flux
scaling factors. That is, for each white dwarf
\begin{equation}
\frac{f_\mathrm{wd}}{F_\mathrm{wd} \times \pi} = \left (\frac{R_\mathrm{wd}}{d} \right)^{2}
\label{e-wddist}
\end{equation}
where  $R_\mathrm{wd}$  is the  white  dwarf  radius  and $d$  is  the
distance, parameters that are both known for each SNIa progenitor. The
model spectra  are provided  in vacuum  wavelengths, which  we convert
into air wavelengths.

The orbital  periods of the  SNIa progenitors in our  four populations
are  short ($\la$80  hours,  especially those  that  merge within  the
Hubble time,  $\la$1.5 hours;  see Figure\,\ref{f-param}).   Hence, we
need  to apply  a wavelength  shift  due to  the corresponding  radial
velocity variation  (shortened by the  inclination factor) to  each WD
synthetic spectrum  component.  Moreover, the  spectrum of each  WD is
affected by the corresponding gravitational redshift.

We use the following equations to get the gravitational redshift $Z$
for each WD (in km/s)
\begin{equation}
Z_1 = 0.635 \left( \frac{M_1}{R_1} + \frac{M_2}{a} \right), \hspace{0.5cm}
Z_2 = 0.635 \left( \frac{M_2}{R_2} + \frac{M_1}{a} \right);
\end{equation}
where the masses ($M_1$, $M_2$) and  radii ($R_1$, $R_1$) are in solar
units and  $a$ is the orbital  separation, also known for  each binary
from  Kepler's third  law and  given in  solar radii.  This expression
takes into account the gravitational potential acting on a WD owing to
its  WD  companion.  We   convert  the  gravitational  redshifts  into
wavelength shifts that we then apply to each WD synthetic spectrum.

The maximum radial velocity shift $K_1$ for WD$_{1}$ is obtained
following
\begin{equation}
K_\mathrm{1} = \left[ \frac{2\pi G\, (M_{2}\sin i)^3}{\Porb\, (M_{1}+M_{2})^2} \right]^{1/3}
\end{equation}
with $i$ the  orbital inclination, \Porb\, the orbital  period and $G$
the gravitational constant. We then obtain the maximum radial velocity
shift $K_2$ for WD$_{2}$ as  $K_1 \frac{M_1}{M_2}$. The maximum radial
velocity shifts  are converted into  wavelength shifts and  applied to
the WD  synthetic spectra. We assume  a zero systemic velocity  in all
cases.

We finally obtain the  double-degenerate (combined) spectrum by adding
the  observed fluxes  of  each  WD component,  corrected  both by  the
gravitational  and maximum  radial velocity  shifts. Two  examples are
shown in  Figure\,\ref{f-DD}, where we  also display a zoom-in  to the
spectra  around the  H$\alpha$  line region.   The  H$\alpha$ line  is
typically     used    for     identifying    double-lined     binaries
\citep[e.g.][]{Koester01}.  Double-lined  binaries allow  sampling the
orbital  motion of  the two  stars, hence  one can  derive the  radial
velocity semi-amplitudes and the mass  ratio, which together with some
elaborated further  analysis allows  deriving the component  masses of
the  WDs (see  for  example  \citealt{Maxted02,Rebassa17}).  With  the
component masses and  the orbital periods at hand one  can then easily
evaluate whether or  not the binary will merge within  the Hubble time
and explode as a SNIa and/or  a sub-Chandrasekhar SNIa, or simply form
a massive WD.

Figure\,\ref{f-DD} shows two WD synthetic  spectra from our Ch. direct
population. Both display double-lined profiles  from which we would be
able to measure the orbital periods  and component WD masses and hence
identify such system as a SNIa progenitors.

\begin{figure*}
    \includegraphics[angle=-90,width=\columnwidth]{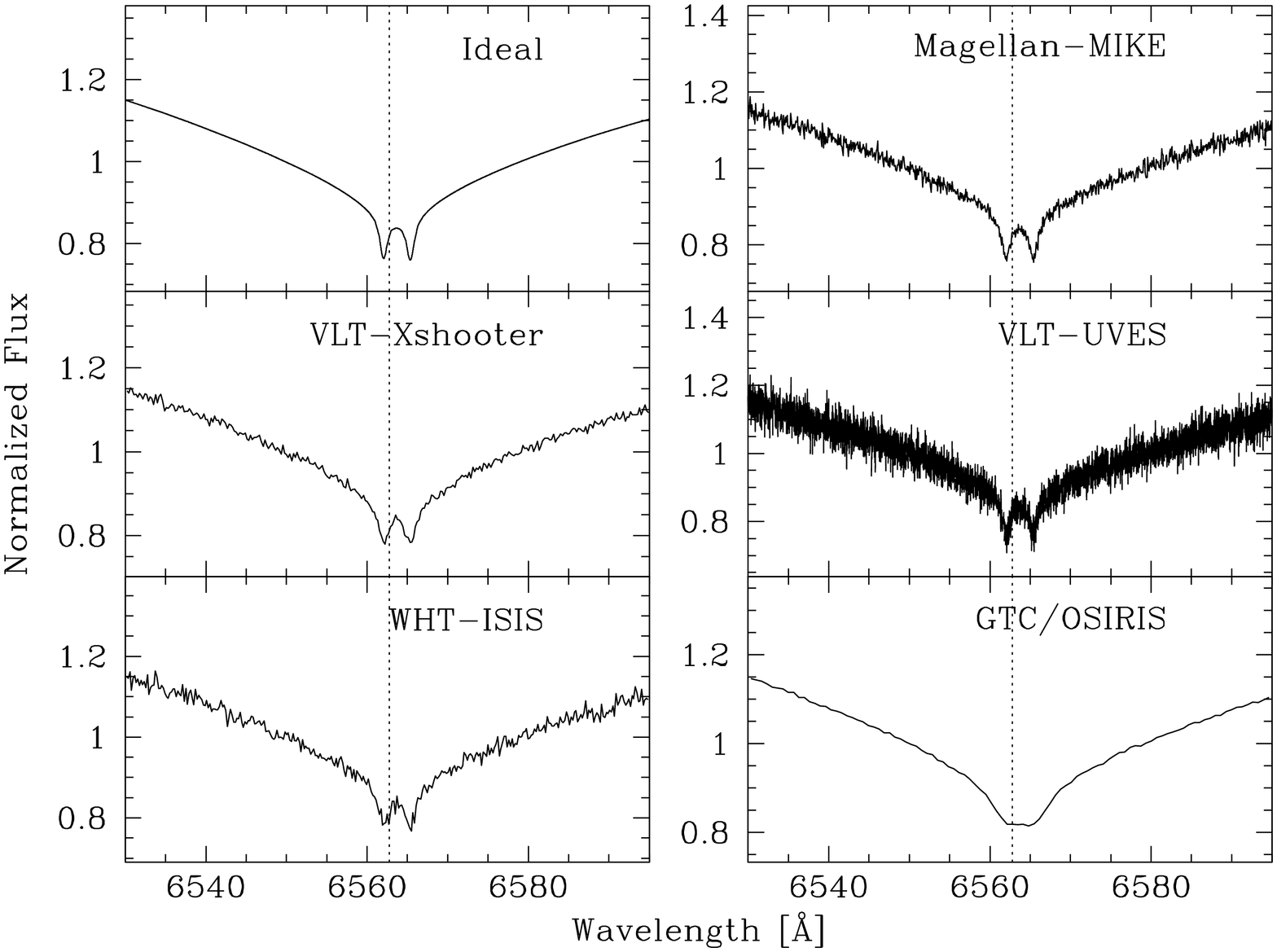}
    \includegraphics[angle=-90,width=\columnwidth]
{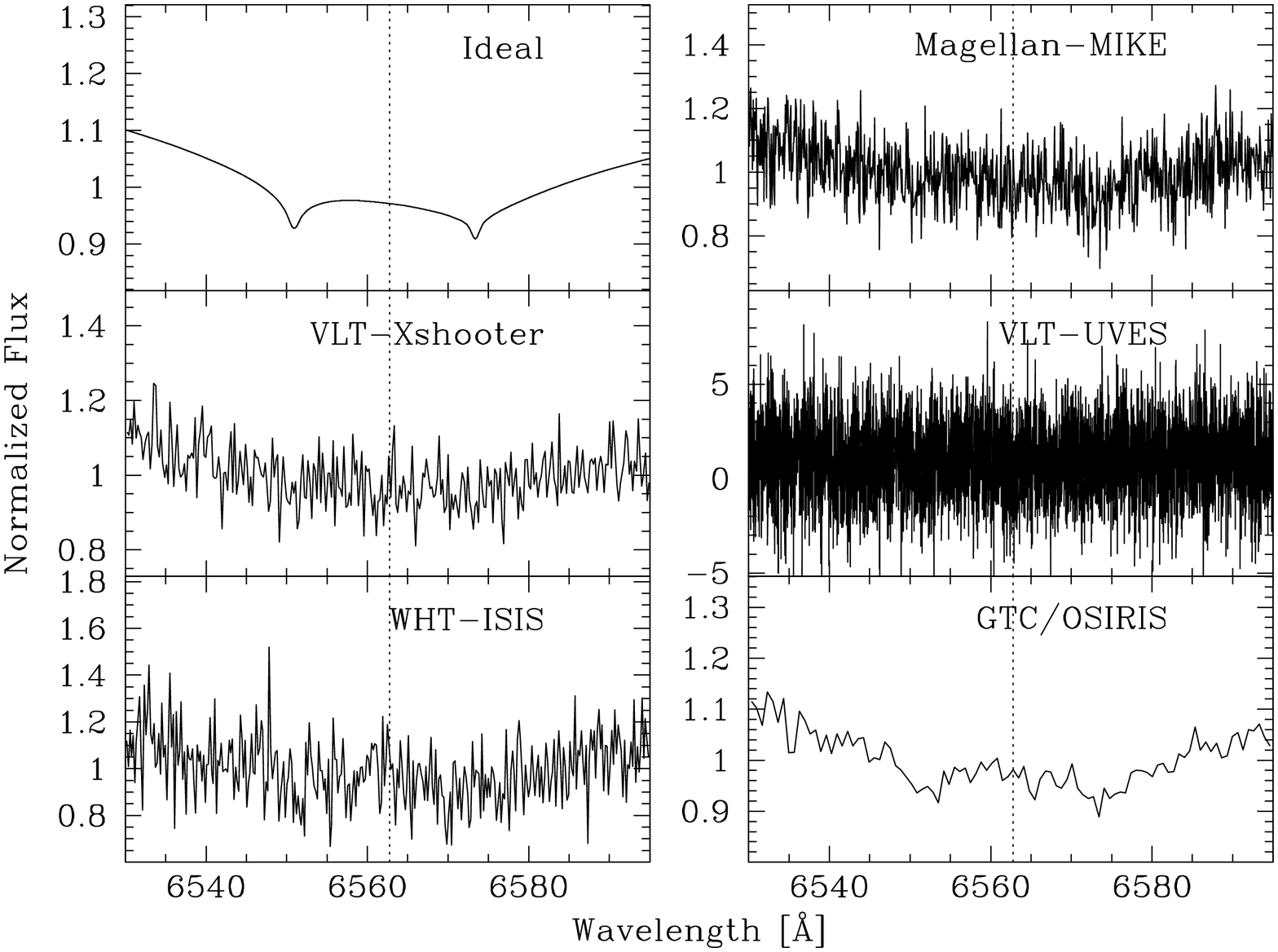}
    \includegraphics[angle=-90,width=\columnwidth]
{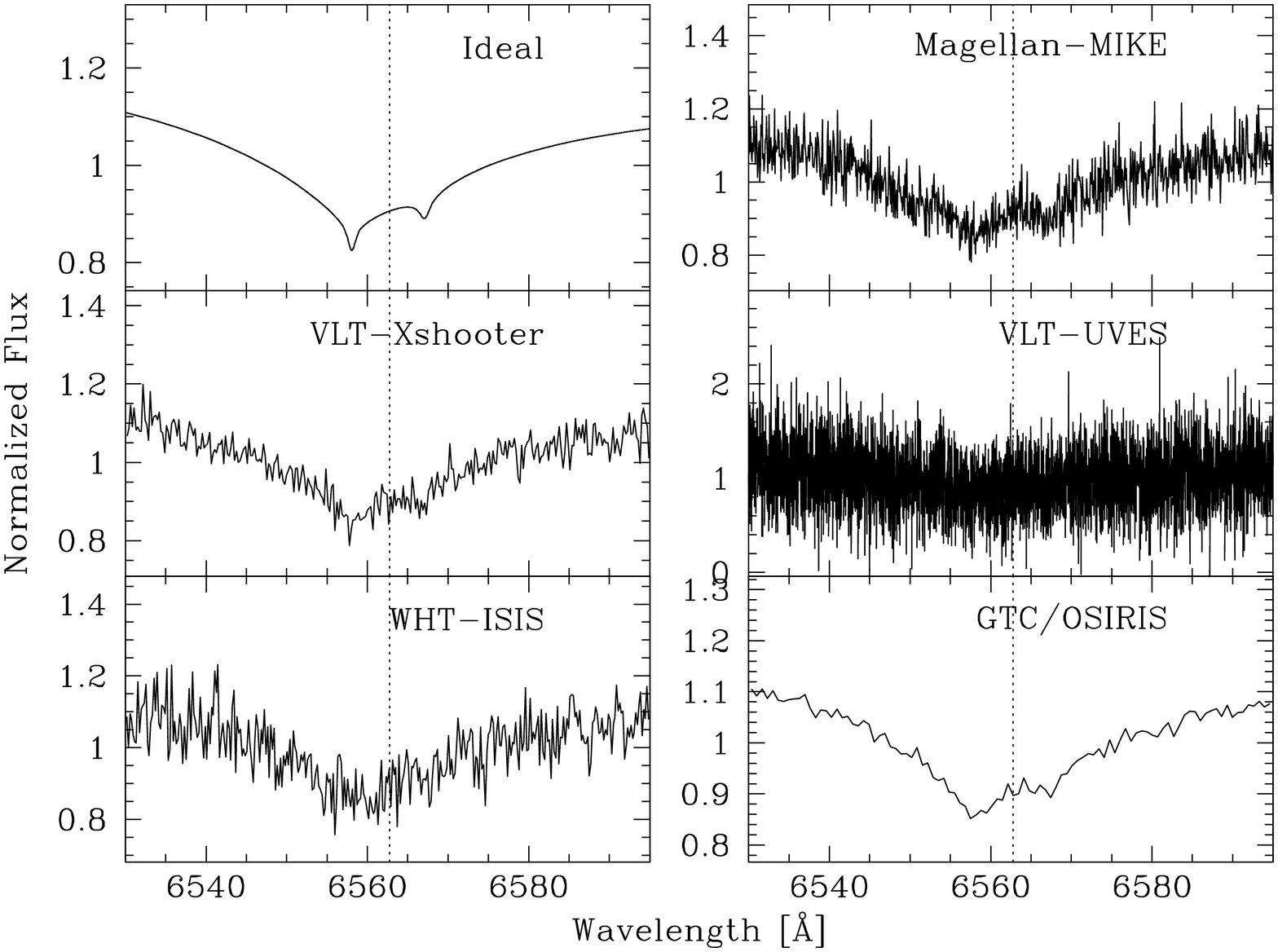}
    \includegraphics[angle=-90,width=\columnwidth]
{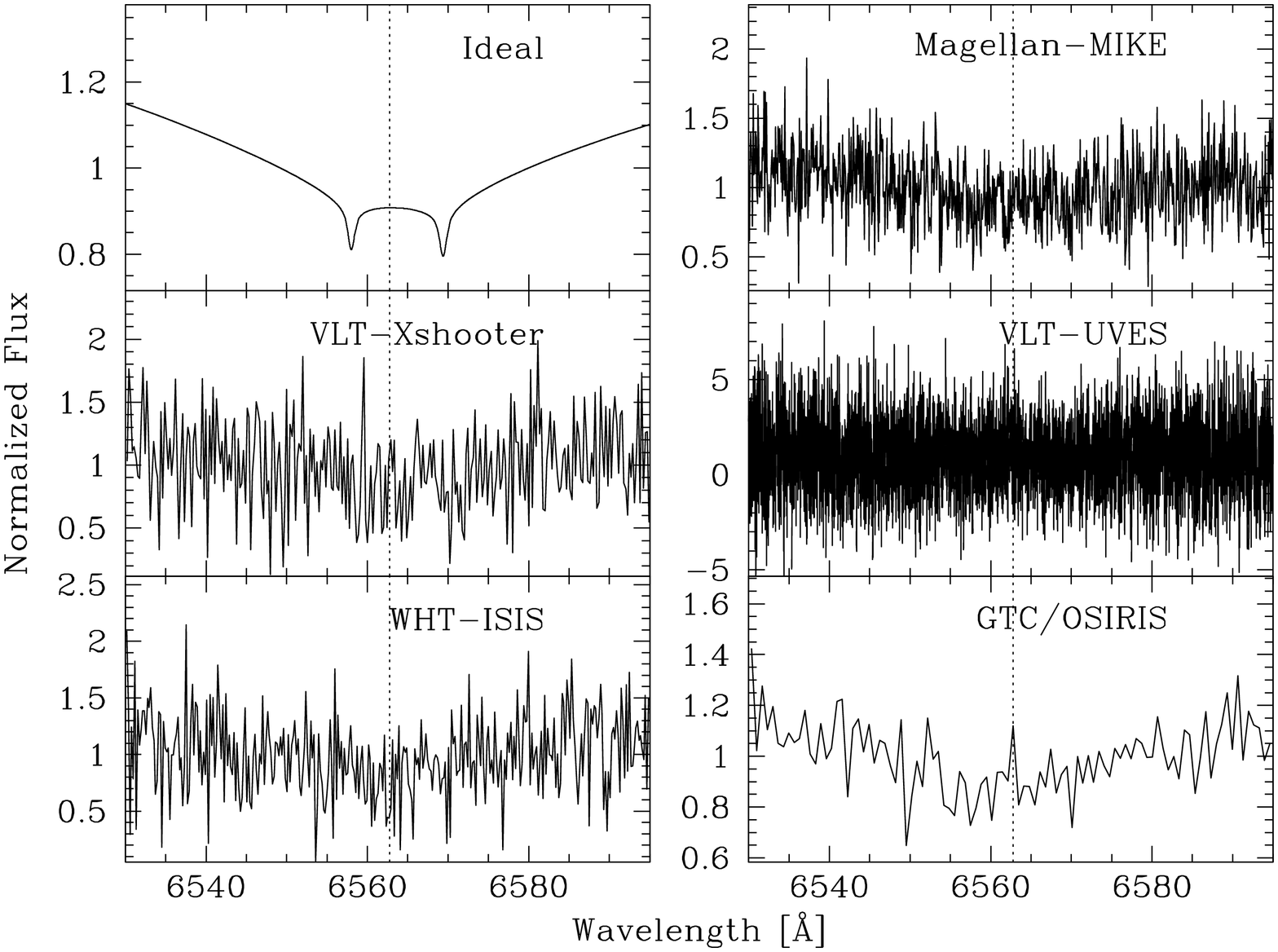}
    \caption{A  zoom-in  to  the  H$\alpha$ region  of  the  synthetic
      spectra of  four Ch.  direct  SNIa progenitors (i.e.  double WDs
      that  merge  within  the  Hubble  time and  with  a  total  mass
      exceeding  1.3\,\Msun)   as  observed  by  the   five  different
      telescope/spectrograph configurations  considered in  this work.
      For comparison, we also show the ``ideal'' spectra, i.e. spectra
      not affect by any observational bias, of the four binaries.  The
      temperatures   and    masses   of   the   WD    components   are
      $M_1=0.83\,$M$_{\odot}$,                $M_2=0.68\,$M$_{\odot}$,
      $T_1=15778$\,K,     $T_2=15472$\,K     (top    left     panels);
      $M_1=1.13\,$M$_{\odot}$,                $M_2=0.68\,$M$_{\odot}$,
      $T_1=35076$\,K,     $T_2=22896$\,K    (top     right    panels);
      $M_1=1.10\,$M$_{\odot}$,                $M_2=0.39\,$M$_{\odot}$,
      $T_1=19234$\,K,   $T_2=33677$\,K   (bottom  left   panels)   and
      $M_1=0.81\,$M$_{\odot}$,                $M_2=0.76\,$M$_{\odot}$,
      $T_1=15107$\,K, $T_2=14533$\,K (bottom right panels)}
    \label{f-spec}
\end{figure*}

\section{Observational effects}
\label{s-obs}

The  double WD  synthetic  spectra obtained  in  the previous  section
represent ideal spectra in the sense  that they are given at virtually
infinity signal to noise ratio (SNR)  as well as at a resolution which
is   typically   larger   than    the   ones   provided   by   current
spectrographs. Therefore, in order to  evaluate whether the double WDs
would  be  clearly  detected  as  double-lined  binaries,  we  require
incorporating  observational effects  in the  synthetic spectra,  i.e.
adding artificial  noise and  downgrading the spectral  resolution. To
that end  we evaluate how the  synthetic spectra of our  four selected
SNIa  progenitor populations  would look  like if  these objects  were
observed  by the  following  telescopes/spectrographs:  the 8.2m  Very
Large   Telescope   (VLT)   equipped  with   the   UVES   spectrograph
($R=$110,000), the  VLT equipped with X-Shooter  ($R=7,450$), the 4.2m
William Herschel  Telescope (WHT) equipped with  ISIS ($R=8,350$), the
6.5m  Magellan  Clay telescope  equipped  with  the MIKE  spectrograph
($R=22,000$)  and the  10.4m Gran  Telescopio Canarias  (GTC) equipped
with OSIRIS  ($R=2500$). The choice of  these telescopes/spectrographs
was made with the aim of  covering a wide range of telescope apertures
(which translate into different SNR for the same spectrum assuming the
same exposure time) as well as spectral resolutions.

It becomes  clear from Figure\,\ref{f-param} that  the orbital periods
of the  direct SNIa progenitors (both  the Ch. direct and  SCh. direct
populations) are  very short ($\la  1.5$ hours), independently  of the
common  envelope formalism  adopted. This  implies the  exposure times
need to  be short  if we  were to observe  such systems,  otherwise we
would not sample enough points of the radial velocity curves and, more
importantly, we would not be able  to distinguish the double lines due
to orbital  smearing. We thus assume  an exposure time of  10 minutes,
which is a good comprise to  avoid orbital smearing and to have enough
radial velocities sampling the orbital phases.

Thus fixing a 10 minute exposure  time, we determined the expected SNR
as   a  function   of  $g$   magnitude  for   each  of   our  selected
telescopes/instruments. We  did this  by making  use of  the available
exposure time  calculators for each telescope/instrument  pair. In all
cases  we assumed\footnote{Observing  conditions can  be specified  in
  service  mode observations,  however since  we are  also considering
  telescopes for  which only  visitor mode is  possible we  decided to
  adopt a  typical average seeing  of 1'', and optimal  conditions for
  observing relatively faint objects, i.e.  gray time and a relatively
  high air  mass. We note  that modifying the observing  conditions to
  better ones  does not  considerably affect  the results  obtained in
  this work.} a  moon phase of 0.5  (or gray time), an  airmass of 1.5
and a seeing of 1''. We fitted third order polynomials to the obtained
SNR   versus    magnitude   relations,   which   we    illustrate   in
Figure\,\ref{f-snr}  (red  solid  lines).   From  these  equations  we
estimated the  SNR of all  synthetic spectra.  From these  values, and
assuming a Gaussian noise distribution, we were able to add artificial
noise  to  the  synthetic  spectra.  Before  that,  the  spectra  were
downgraded to the required spectral resolving power.

\begin{table*}
\caption{Number  of   systems  that   would  be  identified   as  SNIa
  progenitors for  the four  populations considered  in this  work. We
  provide the  numbers for each combination  of telescope/spectrograph
  and CE envelope formalism adopted.}
\begin{center}
\begin{tabular}{ccccccc}
\hline
\hline
  Population & CE formalism & GTC/OSIRIS & Mag./MIKE & VLT/UVES & WHT/ISIS & VLT/X-Shooter \\ 
\hline
 Ch. direct  & $\alpha\alpha$  & 3  &  5 &  1 &  1 & 2 \\
             & $\gamma\alpha$  & 0  &  3 &  0 &  1 & 1 \\
 SCh. direct & $\alpha\alpha$  & 3  &  2 &  0 &  0 & 1 \\
             & $\gamma\alpha$  & 0  &  0 &  0 &  0 & 0 \\
 Ch. nmer    & $\alpha\alpha$  & 16 & 77 &  7 & 22 & 47\\
             & $\gamma\alpha$  & 15 & 79 &  6 & 15 & 48\\
 SCh. nmer   & $\alpha\alpha$  & 6  & 25 &  2 &  8 & 19\\ 
             & $\gamma\alpha$  & 3  & 24 &  2 &  6 & 17\\ 
\hline
\end{tabular}
\end{center}
\label{t-numbers}
\end{table*}

As  it  can be  clearly  seen  from Figure\,\ref{f-snr},  only  bright
objects ($g<17$ mag) would achieve a SNR larger than 10 if observed by
the  combinations  of  telescope  aperture  and  spectral  resolutions
considered, except  for the  VLT/UVES pair,  where only  the brightest
targets  ($g\leq  15$ mag)  would  pass  this  cut. This  implies  the
majority  of both  Ch.  and  SCh.   SNIa direct  progenitors would  be
associated  to  rather  low  SNR  spectra  (since  most  objects  have
magnitudes  above $g=18$  mag,  see  Figure\,\ref{f-param}), which  is
expected  to affect  considerably  the detection  of the  double-lined
profiles in the  spectra. This situation changes for the  Ch. nmer and
the  SCh.   nmer  populations,  that  we  recall  include  those  SNIa
progenitors that do not merge within  the Hubble time.  In these cases
the  orbital periods  are considerably  longer (between  10--80 hours;
Figure\,\ref{f-param}), thus  allowing increasing the  exposures times
to more  than 10  minutes since  we would not  be affected  by orbital
smearing.  However, since at the  time of a hypothetical observing run
one  would not  have at  hand any  previous information  regarding the
orbital periods,  we decided to  keep the  exposure times fixed  at 10
minutes for the non-merger populations too.

In Figure\,\ref{f-spec}  we show the  synthetic spectra zoomed  to the
H$\alpha$ region for the five telescope/instrument pairs considered of
four direct SNIa progenitors.  As we have already mentioned, H$\alpha$
is a widely common spectral feature used to both identify double-lined
binaries  and to  measure  the orbital  periods  and component  masses
\citep{Koester01,     Maxted02,     Rebassa17}.      Inspection     of
Figure\,\ref{f-spec} reveals a wide variety of different possibilities
for  the  clear  identification  of the  double-lined  profiles.   For
instance, these can be easily identified in the spectra illustrated in
the  top  left  panels  in  all  cases  except  when  considering  the
GTC/OSIRIS configuration,  where the low  resolution is not  enough to
clearly  resolve  the  two  absorption  lines  despite  the  high  SNR
achieved. Conversely, only when considering the GTC/OSIRIS pair we can
clearly identify the profiles  when inspecting the spectra illustrated
in  the top-right  panels.   In the  bottom-left  panels, the  spectra
resulting   from  the   Magellan/MIKE,  VLT/XShooter   and  GTC/OSIRIS
configurations reveal the two  absorption profiles for this particular
WD binary, whilst no double absorption profiles can be detected in any
of the spectra displayed in  the bottom-right panels.  This implies we
would  not be  able to  measure  the WD  masses for  this system  and,
consequently, we would not detect it as a SNIa progenitor.

In the next Section we analyse in detail how the observational effects
here  described  affect  the  detectability  of  the  SNIa  progenitor
population as a whole.

\section{Results}
\label{s-results}

In order to evaluate the  impact of the observational effects described
in the  previous section in  the detection of double-lined  profile WD
binaries  we  provide  in  Table\,\ref{t-numbers}  the  number  of  WD
binaries  that would  be able  to  be identified  as SNIa  progenitors
(based  on  the  clear  identification  of the  two  profiles  in  the
H$\alpha$ region)  for the four  populations considered in  this work,
taking into  account both the  CE formalism adopted and  the different
telescope/spectrograph configurations.  Table\,\ref{t-numbers} reveals
that the number of identified progenitors  does not depend much on the
CE formalism,  being the only  difference the fact that,  generally, a
slightly less  number of  systems is  identified from  the populations
evolving through  $\gamma\alpha$ CEs. It  also becomes clear  that the
number of identified progenitors  varies considerably depending on the
telescope/spectrograph     configuration,     as     expected     from
Figure\,\ref{f-spec}.

Independently   of  the   CE   formalism  and   telescope/spectrograph
configuration, Table\,\ref{t-numbers}  also shows  that the  number of
identified  SNIa progenitors  is very  low  as compared  to the  total
number     of     progenitor     systems    in     the     populations
(Table\,\ref{t-samples}). If  we consider  the Magellan/MIKE  pair and
the $\alpha\alpha$ synthetic populations, which results in the maximum
number  of SNIa  progenitors identified,  then the  fractions of  SNIa
progenitors  that  are expected  to  be  identified  are 3\%  for  the
Ch. direct population,  4\% for the SCh. direct  population, 0.5\% for
the Ch. nmer population and 0.3\% for the SCh. nmer population. Taking
into  account that  the  complete $\alpha\alpha$  WD binary  synthetic
population contains $\sim$370,000 objects,  of which $\sim$237,000 are
unresolved\footnote{We consider  a synthetic  binary to  be unresolved
  when its  separation on the sky  is less than 1\,\arcsec,  where the
  separation is  calculated following  Eq.12 of  \citet{Too17}.}, then
the  estimated   probabilities  for   finding  SNIa   progenitors  are
($2.1\pm1.0)\times10^{-5}$       (Ch.        direct       population),
$(0.8\pm0.4)\times10^{-5}$       (SCh.        direct       population)
$(3.2\pm0.4)\times10^{-4}$     (Ch.       nmer     population)     and
$(1.1\pm0.2)\times10^{-4}$ (SCh.  nmer  population). The uncertainties
in  the probabilities  are  obtained assuming  Poisson  errors in  the
values provided  in Table\,\ref{t-numbers}.  We obtain  similar values
when  considering  the   $\gamma\alpha$  synthetic  populations.   The
probabilities are lower  for the SCh. populations  since these objects
are a sub-sample of the Ch.  populations (see Figure\,\ref{f-param2}).
Indeed, all the  identified SNIa progenitors in  the SCh.  populations
are also included in the Ch.  direct populations.

\begin{figure}
    \includegraphics[width=\columnwidth]{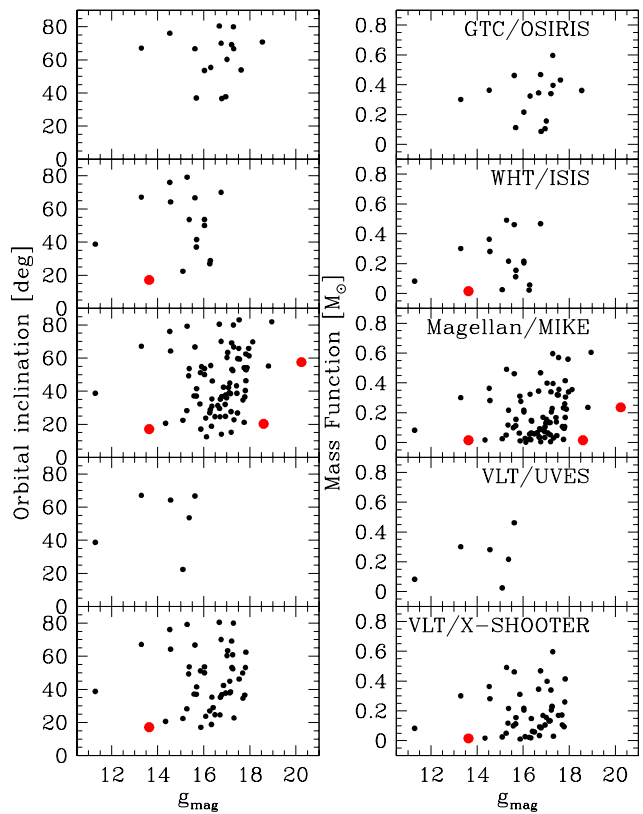}
    \caption{Left  panels: orbital  inclination as  a function  of $g$
      magnitude for all  the SNIa progenitors for  which the simulated
      spectra clearly  display double-lined profiles. Large  red solid
      dots  indicate   direct  SNIa  progenitors,  black   solid  dots
      non-merger  progenitors within  the Hubble  time. Right  panels:
      mass  function as  a  function  of $g$  magnitude  for the  same
      systems.   The  corresponding telescope/spectrograph  pairs  are
      indicated in the top right of each panel.}
    \label{f-inspec}
\end{figure}

Judging  from  Table\,\ref{t-numbers},  the most  efficient  telescope
aperture/resolution combination  seems to be  the one provided  by the
Magellan/MIKE pair, followed by the  VLT/X-Shooter. In both cases, the
apertures are  large enough for  achieving higher SNR spectra  and the
resolving  powers  are  high  enough  for  sampling  the  double-lined
profiles.  This  is also  true for  the WHT/ISIS  configuration, which
results in a similar resolving power  as the one by the VLT/X-Shooter,
but  for a  lower  number  of systems  due  to  the smaller  telescope
aperture. The  GTC/OSIRIS pair achieves  the highest SNR,  however the
spectral  resolution  is rather  low  in  this  case, thus  making  it
difficult to  sample the  two absorption  profiles and  hence reducing
considerably the number of  identified progenitors.  The VLT/UVES pair
is the less efficient  configuration for identifying SNIa progenitors.
This  is due  to the  extremely high  resolving power  achieved, which
limits considerably  the SNR  of the obtained  spectra.  All  this can
clearly be seen in the left panels of Figure\,\ref{f-inspec}, where we
illustrate the orbital  inclination of the binaries  that clearly show
double lines  in their spectra as  a function of their  $g$ magnitudes
for   the   five    telescope   aperture/spectrograph   configurations
considered.  Since the  number of potential progenitors  that are able
to be identified does not dramatically depend on the CE formalism used
(Table\,\ref{t-numbers}),   we   choose    for   this   exercise   the
$\gamma\alpha$ samples, since the  ensemble properties of the binaries
resulting from these simulations better  agree with those derived from
observations \citep{Nel00, Too12}.

From Figure\,\ref{f-inspec} (left panels)  it becomes obvious that, as
expected, larger aperture telescopes are more suitable for identifying
the double-lined  profiles in  the spectra  of fainter  objects, being
$\sim$19 mag the magnitude limit. This is the case for the GTC/OSIRIS,
Magellan/MIKE and  VLT/X-Shooter configurations.  However, it  is also
clear that,  as we mentioned  before, not only the  telescope aperture
but also the  resolving power of the instrument  affects the magnitude
limit for identifying the double lines in the spectra. For example, in
the case  of the  VLT equipped  with the  UVES instrument,  the double
lines can only be identified for  systems with $g$ magnitudes below 16
mag due to  the extremely high resolving power  achieved (which limits
the SNR). The relatively high  resolving power of the MIKE instrument,
followed  by   the  X-Shooter  spectrograph,  makes   this  the  ideal
instrument among the  larger aperture telescopes for  the detection of
the double lines.

The  orbital  inclination  plays  also   an  important  role  for  the
detectability of the  double lines. As can be seen  in the left panels
of Figure\,\ref{f-inspec}, it is not possible to identify double-lined
systems when the inclinations are  lower than $\sim$20 degrees, simply
because  the two  lines  are  smeared. This  effect  is stronger  when
considering  the   GTC  equipped   with  the  low   resolution  OSIRIS
instrument. In this  case, the orbital inclinations need  to be higher
than $\sim$40 degrees.

In  the right  panels of  Figure\,\ref{f-inspec} we  display the  mass
function versus the  $g$ magnitudes for the same  systems displayed in
the left panels. The mass function is defined as:

\begin{equation}
m  = \frac{K_\mathrm{1}^3  P_\mathrm{orb}}{2\pi G}  = \frac{(M_{2}\sin
  i)^3}{(M_{1}+M_{2})^2},
\end{equation}

\noindent where $M_{1}$ is in this  case the brighter star in a binary
and  $K_{1}$ its  semi-amplitude  velocity.  Since  the mass  function
depends only  on $P_\mathrm{orb}$ and $K_\mathrm{1}$,  values that can
be relatively  easy determined  observationally even  for single-lined
binaries,  this  quantity  may  help  in providing  clues  on  how  to
efficiently target SNIa progenitors. However, as it can be seen in the
right panels of  Figure\,\ref{f-inspec}, there seems to  be no obvious
trend, since  a wide range of  values are possible among  all possible
SNIa progenitors.

We conclude  identifying both direct  and non-merger and both  Ch. and
SCh.  SNIa  progenitors  with   our  current  optical  telescopes  and
instrumentation  is  extremely  challenging  due  to  their  intrinsic
faintness.

\section{Discussion}
\label{s-discuss}

The  results  presented in  this  paper  demonstrate that  identifying
double-degenerate   SNIa  progenitors   is  extremely   hard  due   to
observational biases. In other words,  the probability for detecting a
double WD  SNIa progenitor  in the  Galaxy based  on the  detection of
double-lined absorption profiles in the  spectrum is very low with our
current instrumentation, since the vast  majority of these systems are
intrinsically faint.  Increasing the current size of known WDs e.g. by
analysing  the recent  superb  sample of  $\sim$8500 \emph{Gaia}  data
release 2 WDs within 100pc  from the Sun \citep{Jimenez2018}, does not
seem to  be the solution  given that possible SNIa  progenitors within
this sample are also expected to be faint. In the following we analyse
possible ways for increasing the  probability of detection and discuss
alternative ways for finding SNIa double WD progenitors.

\subsection{The next generation of large-aperture telescopes}

A  clear way  to  move forward  includes  improving our  observational
facilities.   Fortunately,  the   next  generation  of  large-aperture
($\simeq30$m) optical telescopes such  as the European Extremely Large
Telescope   \citep[E-ELT;][]{EELT},  the   Great  Magellan   Telescope
\citep[GMT;][]{GMT} or the  Thirty Meter Telescope \citep[TMT;][]{TMT}
will allow observing down to deeper magnitudes at a much lower cost in
terms of exposure times. This  is expected to increase the probability
of detecting  double WD SNIa  progenitors. For instance, if  we assume
these telescopes  to achieve  a reasonably  high SNR  for a  10 minute
exposure for objects down to 23 magnitudes (i.e. we are not limited by
the  noise in  the  spectrum but  on the  orbital  inclination of  the
systems for detecting the double-lined profiles), then the probability
for finding e.g.  Ch.  direct  SNIa progenitors increases one order of
magnitude          from         $(2.1\pm1.0)\times10^{-5}$          to
$(3.9\pm0.4)\times10^{-4}$.

\subsection{Uncertainties in our numerical simulations}

It is possible that our numerical  simulations predict a low number of
SNIa  progenitors,  in which  case  we  would be  underestimating  the
probability of their detection. The observed SNIa rate integrated over
a   Hubble   time   is   $(13\pm   1)\times   10^{-4}$M$_{\odot}^{-1}$
\citep[][and references  therein]{Maoz2017b}.  On  the other  hand the
integrated  rates  in our  simulations  for  Chandrasekhar mergers  of
double white dwarfs are a factor of a few lower than the observed rate
($(4.2-5.5)\times 10^{-4}$M$_{\odot}^{-1}$), which,  however, does not
significantly affect the detection probabilities.

It is important to emphasize that  the evolution of double WD binaries
is  not well  understood  yet  and that  our  adopted  modeling of  CE
evolution  (both the  $\alpha$  and $\gamma$  formalisms)  may not  be
adequate  \citep{Woo12,  Pas12, Ge15}.   A  better  treatment of  mass
transfer   could    help   in   increasing   the    number   of   SNIa
progenitors. Hence, exploring population  synthesis models including a
phase of stable non-conservative mass transfer (unfortunately, not yet
implemented in any  BPS code) rather than a first  CE phase seems them
to  be a  worthwhile  exercise.   An additional  factor  to take  into
account is that our results are  based on the outcome of two synthetic
double WD binary  populations that differ only with respect  to the CE
phase.   Uncertainties  on  other  physical  processes,  such  as  the
stability or  mass accretion efficiency  of mass transfer,  affect the
double white dwarf population as well, but to a lesser degree.

\begin{figure*}
        \centering
	 \includegraphics[width=0.49\textwidth]{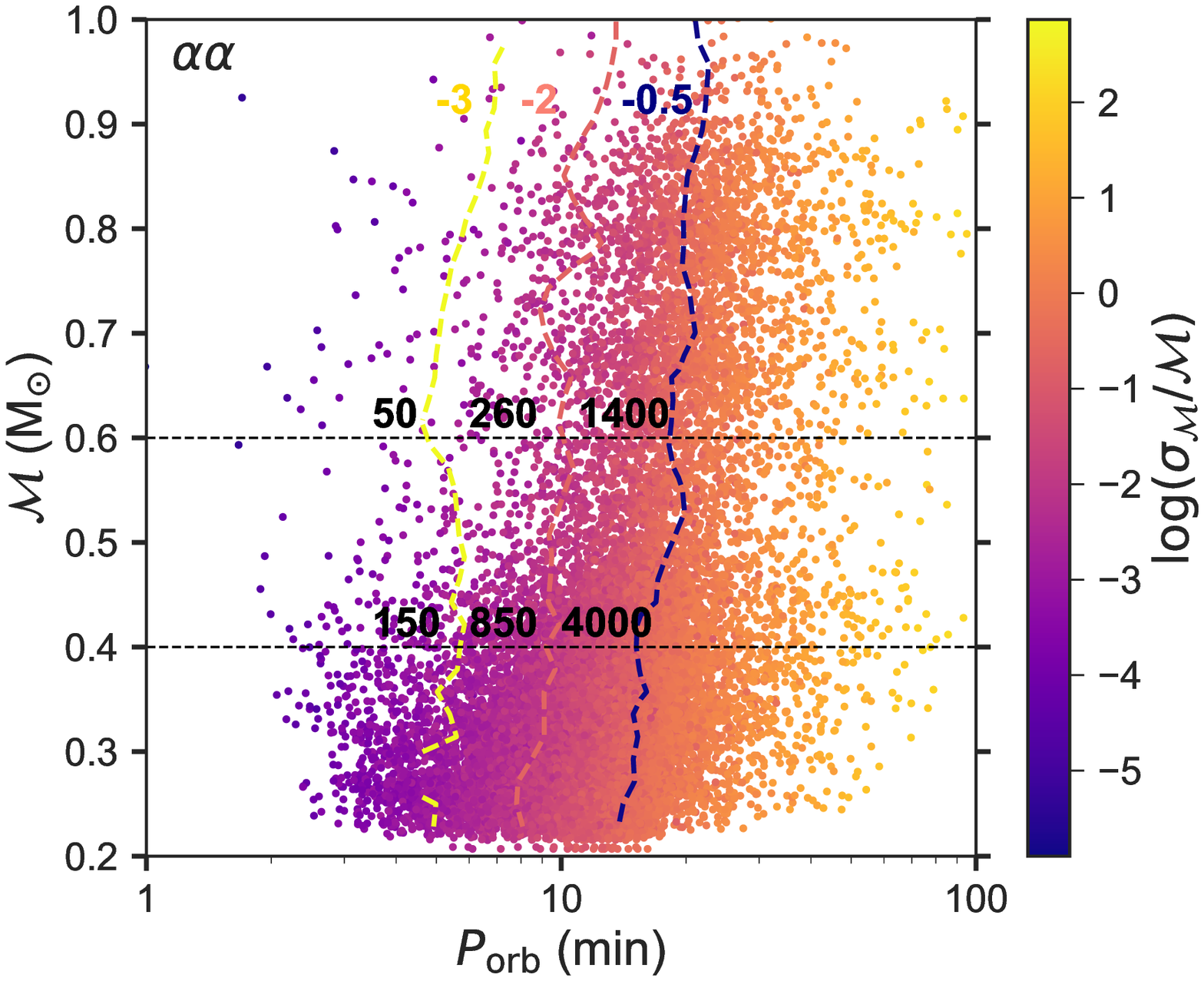} 
	 \includegraphics[width=0.49\textwidth]{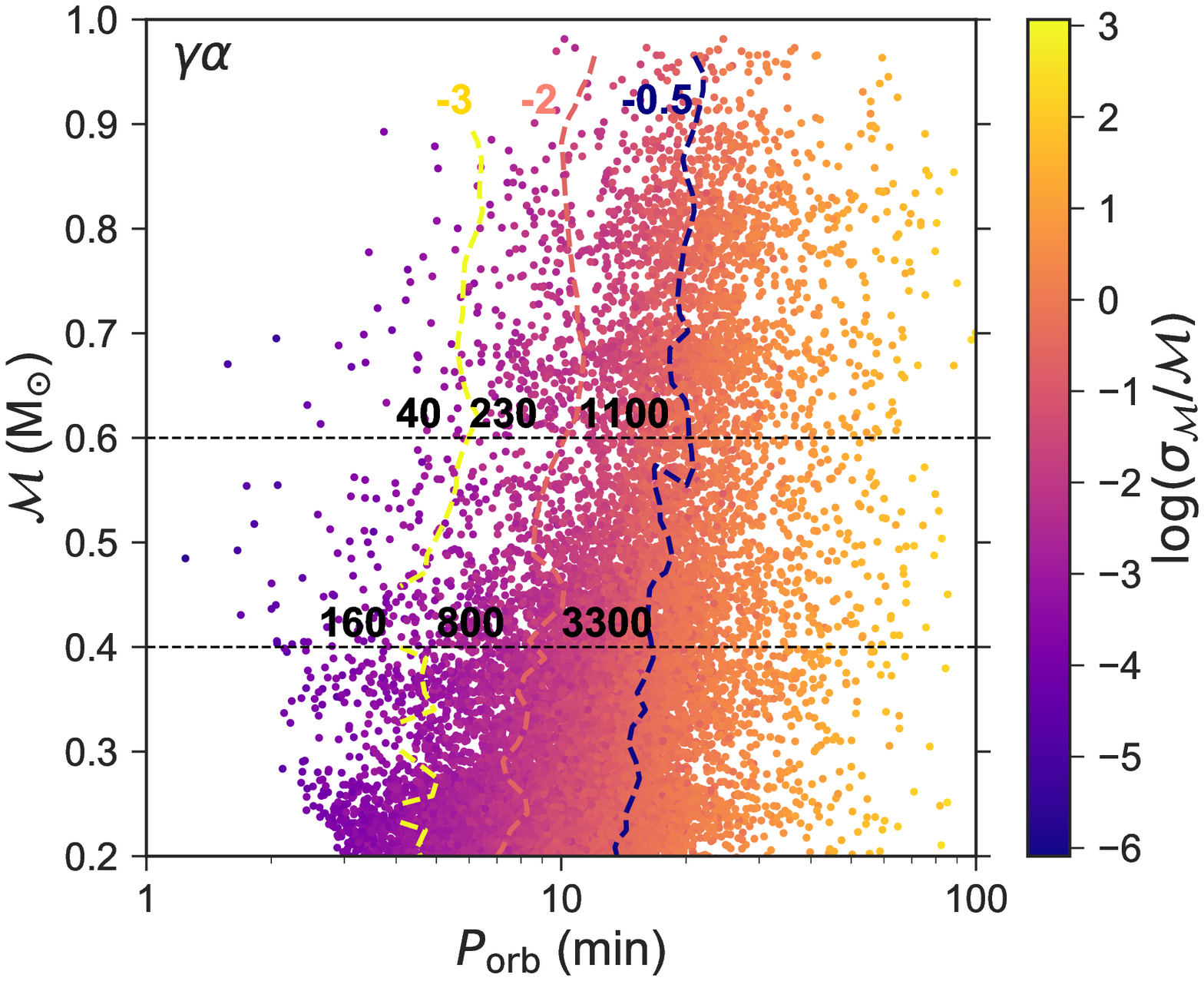}     
         \caption{Relative  error  on  the chirp  mass  $\sigma_{{\cal
               M}}/{\cal  M}$ in  period-chirp  mass parameter  space.
           The dashed  vertical lines are  iso-relative-error contours
           for $\log$($\sigma_{{\cal M}}/{\cal M}$)  $ =$ -3 (yellow),
           -2  (pink) and  -0.5 (blue),  equivalent to  $\sigma_{{\cal
               M}}/{\cal  M}  =  0.1\%$  (yellow),  $1\%$  (pink)  and
           $\sim32\%$  (blue).  The  values on  top of  the horizontal
           lines indicate the number of  systems with a given relative
           error  limit  (i.e.  systems  to  the  left of  the  dashed
           vertical lines)  and with a  chirp mass higher than  0.4 or
           0.6 M$_{\odot}$ (systems above the horizontal lines).}
    \label{fig:LISA}
\end{figure*}

We  also need  to bear  in  mind that  our synthetic  double WD  space
density may  be underestimated. Recently, our  synthetic space density
values were verified by \cite{Too17}  with a comparison between common
synthetic WD  systems based on the  same BPS set-up employed  here and
the nearly-complete  volume-limited sample  of WDs  within 20  pc. The
space  density  of single  WDs  and  resolved  WD plus  main  sequence
binaries (which  represent the most  common WD systems)  are correctly
reproduced within a  factor 2. The synthetic models of  double WDs are
in good  agreement with  their observed number  in 20pc,  albeit given
their  current  small  number  statistics   in  this  sample  (1  +  4
candidates). Another constraint on the space density of double WDs can
be made by  studying the ratio $f$  of double WDs to  single WDs. From
the  above-mentioned  20pc  sample,  $f=0.008-0.04$  for  (unresolved)
double WDs,  whereas our  model $\alpha\alpha$ gives  $f\approx 0.02$,
and model  $\gamma\alpha$ $f\approx 0.04$  \citep[see][]{Too17}. Based
on  radial   velocity  measurements   of  a  sample   of  46   DA  WDs
\cite{MaxtedMarsh99} deduce $f=0.017-0.19$ with a 95\% probability for
periods of hours  to days. From a statistical approach  of the maximum
radial   velocity    measurement   of   $\sim$4000   WDs    in   SDSS,
\cite{BadenesMaoz12} deduce  $f =  0.03-0.20$ for orbits  smaller than
0.05AU.   With a  similar approach  on 439  WDs from  the SPY  survey,
\citet{Maoz2017} find  $f = 0.103$ with  a random error of  $\pm 0.02$
and a  systematic error of $\pm0.015$  for orbits within 4AU.   From a
joint     likelihood     analysis     of    these     two     samples,
$f=0.095\pm0.020$\,(random)$\pm0.010$\,(systematic)  \citep{Maoz2018}.
These  measurements  may imply  that  our  synthetic double  WD  space
densities are  underestimating the  true space density,  but not  by a
factor  more  than  2-5,  which  does  not  significantly  change  our
calculated  value  of  the  probability  of  finding  double  WD  SNIa
progenitors.

\subsection{Eclipsing double WDs}

It is important to emphasise that  we have considered the detection of
SNIa   progenitors  based   only  on   the  clear   identification  of
double-lined  profiles,  which  allows   measuring  the  WD  component
masses. An additional  way of measuring the  masses involves analysing
the        light       curves        of       eclipsing        systems
\citep[e.g.][]{Parsons2011}. Since  the orbital inclinations  and mass
ratios can be relatively well constrained in these cases, by measuring
the  orbital  periods  and  deriving   the  masses  of  the  brightest
components  (e.g.   by  fitting   the  observed  spectrum  with  model
atmosphere spectra), one can then  derive precise values of the masses
of the  two WDs. However,  deriving the semi-amplitude velocity  of at
least one of the WD  components is required for accurately determining
the WD  masses. Thus, so far  only 7 eclipsing double  WD binaries are
known for which the WD masses have been accurately determined, none of
them     being    direct     SNIa    progenitors:     SDSS\,J0651+2844
\citep{Hermes2012}, GALEX\,J1717+6757  \citep{Hermes2014}, NLTT\,11748
\citep{Kaplan2014},  SDSS\,J0751-0141   \citep{Kilic2014},  CSS\,41177
\citep{Bours2015},    SDSS\,J1152+0248    \citep{Hallakoun2016}    and
SDSS\,J0822+3048 \citep{Brown2017}.   It is important to  keep in mind
however  that the  identification of  a large  number of  eclipsing WD
binaries  is  expected  from  the forthcoming  Large  Synoptic  Survey
Telescope \citep[LSST;][]{LSST}. \citet{Kor17} predicted the number of
eclipsing WD binaries LSST will identify  is close to one thousand. It
has to be emphasised that the  authors of that paper employed the same
numerical simulation  code than us,  which easily allows us  to obtain
synthetic spectra for their eclipsing systems and to thus evaluate how
many of them are potential SNIa  progenitors and for how many we could
derive the semi-amplitude velocities of at least one WD component with
our adopted  telescopes/spectrographs.  From the  $\sim$1000 eclipsing
double WDs that  LSST is expected to identify, only  3--7 are found to
be direct  Ch.  SNIa  progenitors depending on  the CE  formalism used
(note that no SCh.  progenitors  are in the samples) and unfortunately
none  of  them  would  be   suitable  for  radial  velocity  follow-up
observations due to their intrinsic faintness.

\subsection{Gravitational waves and LISA}

An  alternative way  of detecting  SNIa progenitors  is by  exploiting
their  gravitational wave  (GW)  radiation.  Double  WD binaries  with
orbital periods  from a  few minutes  to one hour  are expected  to be
detected  through  GW  radiation  by the  Laser  Interferometer  Space
Antenna   \citep[LISA;][]{LISA2017}\footnote{The   LISA  mission   was
  officially approved by ESA in 2017 and scheduled for launch in early
  2030.}.   Using  the  same   population  synthesis  code  and  model
assumptions  as  in this  paper,  \citet{Kor18}  showed that  LISA  is
expected to  individually resolve $>  10^5$ double WD  binaries across
the Milky Way. Here we investigate how many of the LISA detections are
expected to be SNIa progenitors.
 
Long timescales on which double WDs evolve (typically $\sim$Myr) imply
that LISA will catch them in  the inspiral phase.  During inspiral the
evolution of  the GW  signal depends  on the  so-called chirp  mass, a
particular combination of the individual  WD masses, defined as ${\cal
  M} =  (M_1 M_2)^{3/5} (M_1+M_2)^{-1/5}$. This  means that individual
masses $M_1$  and $M_2$ are  difficult to  estimate from GW  data and,
typically, this requires additional assumptions. Thus, in this work we
use   the  chirp   mass  to   select  SNIa   progenitors  among   LISA
detections. In particular, we adopt two thresholds: $0.6\,$M$_{\odot}$
and $0.4\,$M$_{\odot}$. The first one  comes from considering a binary
with     equal    mass     components    and     the    total     mass
$M=1.38\,$M$_{\odot}$. The last one  is determined from our catalogues
as   the  minimum   chirp  mass   among   the  binaries   with  $M   >
1.38\,$M$_{\odot}$.  To compute the GW  signal for binaries in the two
mock  catalogues  we  employ  the  Mock  LISA  Data  Challenge  (MLDC)
pipeline, designed for the analysis  of Galactic GW sources \citep[for
  details see][]{Littenberg2011}. We model double WD waveforms using a
set of 9 parameters: GW  amplitude $\cal{A}$, GW frequency $f=2/P_{\rm
  orb}$,  the   frequency  evolution   or  chirp   $\dot{f}$,  orbital
inclination $i$, polarization angle  $\psi$, initial GW phase $\phi_0$
and  binary  coordinates  on  the sky.   We  estimate  the  respective
uncertainties  by  computing  Fisher  Information  Matrix  \citep[FIM,
  e.g.][]{Shah2012}. We adopt the most  recent LISA mission design and
the noise model from  \citet{LISA2017}, i.e. a three-arm configuration
with $2.5 \times 10^6\,$km arm length. Finally, we assume the duration
of the mission to be of 4 yr.

From GW  data the chirp  mass can be  determined by taking  the lowest
order in a post-Newtonian expansion of the waveform's phase, i.e.
\begin{equation} \label{eqn:Mchirp}
{\cal M} = \frac{c^3}{G} \left( \frac{5}{96} \pi^{-8/3} {\dot f}
\right)^{3/5} f^{11/5},
\end{equation}
where $f$ and $\dot{f}$ are direct GW observables, which uncertainties
and correlation  coefficient can be  extracted from the FIM.   We find
1400  (1100) double  WDs  with  ${\cal M}  >  0.6\,$M$_{\odot}$ and  a
relative error  on the  chirp mass  $< 30\%$  for our  $\alpha \alpha$
($\gamma    \alpha$)    catalogues.     Using   the    threshold    of
$0.4\,$M$_{\odot}$   we   find   4000  (3300)   binaries.    In   Fig.
\ref{fig:LISA}  we represent  the  relative error  on  the chirp  mass
$\sigma_{{\cal M}}/{\cal  M}$ for double  WDs detected by LISA  in the
period -  chirp mass parameter  space.  Figure \ref{fig:LISA}  shows a
gradual  decrease  in  $\sigma_{{\cal  M}}/{\cal M}$  from  longer  to
shorter orbital  periods.  This is  because short period  sources have
larger chirps, which  makes it easier to determine the  chirp mass and
its uncertainty.  Furthermore, binaries  with large chirp  masses have
large  GW  amplitudes  (${\cal   A}  \propto  {\cal  M}^{5/3}$),  that
facilitate their detection. These two facts reflects in high number of
SNIa progenitors detected by LISA.

\section{Summary and conclusions}
\label{s-concl}

With the aim of evaluating the observability of double-degenerate SNIa
progenitors we simulated the double WD binary population in the Galaxy
and obtained  synthetic optical spectra  for each progenitor.  To that
end  we  considered a  set  of  ground-based telescopes  of  different
diameter sizes and  equipped with spectrographs covering  a wide range
of spectral resolutions.

We analysed the detectability of clear H$\alpha$ double-lined profiles
in  the synthetic  spectra and  considered a  positive detection  as a
sufficient  condition  for  deriving   accurate  orbital  periods  and
component  masses of  the  two WDs.   In these  cases  we assumed  the
systems would be identified as SNIa progenitors.  Due to the intrinsic
faintness of  the double-degenerate  SNIa population,  our simulations
indicate that only a handful of  objects are expected to be found with
clear  double-lined profiles  in their  spectra, which  resulted in  a
probability    of   finding    double   WD    SNIa   progenitors    of
$(2.1\pm1.0)\times10^{-5}$  (for  the  direct  classical  Chandrasekhar
progenitor population) and  $(0.8\pm0.4)\times10^{-5}$ (for the direct
sub-Chandrasekhar progenitor population).  These results do not depend
significantly on the  formalism of common envelope  adopted.  We found
the  best  combination  of  telescope/spectrograph  for  finding  SNIa
progenitors is the Magellan Clay/MIKE, followed by the VLT/X-Shooter.

Forthcoming  large-aperture telescopes  are expected  to increase  the
probability for finding double WD SNIa progenitors by $\sim$1 order of
magnitude.  Although  this  is  a  considerably  large  increase,  the
probability for finding these  objects remains low ($\sim10^{-4}$). We
also analysed how eclipsing binaries can help in increasing the number
of  identified SNIa  progenitors, and  concluded that,  even with  the
outcome of LSST,  the probability remains unchanged.  Our results thus
clearly show that identifying double-degenerate progenitors of SNIa is
extremely  challenging.  It  is   not  surprising  then  that  current
observational studies  have failed at  finding such systems.  We hence
conclude that the lack of observed double WD SNIa progenitors is not a
sufficient  condition for  disregarding the  double-degenerate channel
nor the sub-Chandrasekhar models for SNIa.

Fortunately, thanks to the new  window of gravitational wave radiation
observations that LISA will open,  the expectations for finding double
WD SNIa progenitors are highly encouraging. Our results show that LISA
should  be  able  to  find  $\ga$1000 SNIa  progenitors  by  means  of
measuring the chirp masses of the  WD binaries, which will allow us to
robustly  confirm or  disprove  (in  the case  of  no detections)  the
relevance of double WD binaries for producing SNIa. It has to be noted
however that follow-up  spectroscopic/photometric observations will be
required  to   measure  the   individual  masses  of   the  identified
progenitors.

\section*{Acknowledgements}

This  work was  supported  by  the MINECO  Ram\'on  y Cajal  programme
RYC-2016-20254, by  the MINECO  grant AYA\-2017-86274-P, by  the AGAUR
(SGR-661/2017), by  the Netherlands  Research Council NWO  (grant VENI
[nr.  639.041.645])  and  byNWO  WARP Program  (grant  NWO  648.003004
APP-GW).

We thank  Detlev Koester for providing  us with his white  dwarf model
atmosphere spectra and Elena Maria Rossi for her suggestions.

%%%%%%%%%%%%%%%%%%%%%%%%%%%%%%%%%%%%%%%%%%%%%%%%%%



%\bibliographystyle{mnras}
%\bibliography{biblio} % if your bibtex file is called example.bib

% Don't change these lines
\bsp	% typesetting comment
\label{lastpage}
\end{document}